\documentclass[12pt]{iopart}

\usepackage{iopams}  
\expandafter\let\csname equation*\endcsname\relax

\expandafter\let\csname endequation*\endcsname\relax
\usepackage{amsmath}
\usepackage{amssymb}
\usepackage{booktabs}
\usepackage{graphicx}
\usepackage{multirow}
\usepackage{array}
\usepackage{booktabs}
\usepackage{hyperref}
\usepackage{tabularx}
\begin{document}

% \title[A Review of Simulation-based Inference on Gravitational Wave Astronomy]%
      % {A Review of Simulation-based Inference on Gravitational Wave Astronomy}
\title[Recent Advances in Simulation-based Inference for Gravitational Wave Data Analysis]
      {Recent Advances in Simulation-based Inference for Gravitational Wave Data Analysis}
      
\author{Bo Liang$^{\dagger}$}
\address{Center for Gravitational Wave Experiment, National Microgravity Laboratory, Institute of Mechanics, Chinese Academy of Sciences, Beijing 100190, China}
\address{Taiji Laboratory for Gravitational Wave Universe (Beijing/Hangzhou), University of Chinese Academy of Sciences (UCAS), Beijing 100049, China}
\vspace{10pt}

\author{He Wang$^{\dagger}$}
\address{International Centre for Theoretical Physics Asia-Pacific (ICTP-AP), University of Chinese Academy of Sciences (UCAS), Beijing 100049, China;}
\address{Taiji Laboratory for Gravitational Wave Universe (Beijing/Hangzhou), University of Chinese Academy of Sciences (UCAS), Beijing 100049, China}
\ead{hewang@ucas.ac.cn}
\vspace{10pt}

\footnotetext{$^{\dagger}$ These authors contributed equally to this work.}

% \begin{indented}
% \item[]August 2017
% \end{indented}

% \begin{abstract}
% The detection of gravitational waves by the LIGO-Virgo-KAGRA collaboration has ushered in a new era of observational astronomy, shifting the focus towards rapid and detailed parameter estimation and population-level analyses. Traditional Bayesian inference methods, such as Markov Chain Monte Carlo (MCMC), face significant computational challenges due to the high-dimensional parameter spaces and complex noise characteristics of gravitational wave data. This review explores the application of simulation-based inference (SBI) methods, particularly those leveraging machine learning techniques like normalizing flows (NF) and neural posterior estimation (NPE), to enhance gravitational wave parameter estimation. We discuss the theoretical foundations of SBI methods, including neural posterior estimation, neural ratio estimation, neural likelihood estimation, flow matching, and consistency models. We also highlight their applications in various gravitational wave data processing scenarios, such as single-source parameter estimation, overlapping signal analysis, testing general relativity, and population studies. These methods offer substantial speed improvements over traditional techniques while maintaining accuracy, making them valuable tools for future gravitational wave astronomy.
% \end{abstract}

\begin{abstract}
The detection of gravitational waves by the LIGO-Virgo-KAGRA collaboration has ushered in a new era of observational astronomy, emphasizing the need for rapid and detailed parameter estimation and population-level analyses. Traditional Bayesian inference methods, particularly Markov chain Monte Carlo, face significant computational challenges when dealing with the high-dimensional parameter spaces and complex noise characteristics inherent in gravitational wave data. This review examines the emerging role of simulation-based inference methods in gravitational wave astronomy, with a focus on approaches that leverage machine-learning techniques such as normalizing flows and neural posterior estimation. We provide a comprehensive overview of the theoretical foundations underlying various simulation-based inference methods, including neural posterior estimation, neural ratio estimation, neural likelihood estimation, flow matching, and consistency models. We explore the applications of these methods across diverse gravitational wave data processing scenarios, from single-source parameter estimation and overlapping signal analysis to testing general relativity and conducting population studies. Although these techniques demonstrate speed improvements over traditional methods in controlled studies, their model-dependent nature and sensitivity to prior assumptions are barriers to their widespread adoption. Their accuracy, which is similar to that of conventional methods, requires further validation across broader parameter spaces and noise conditions.

\end{abstract}

%\maketitle
\ioptwocol

%------------------------------------------------------------------------------
\section{Introduction}
%------------------------------------------------------------------------------
% DONE: http://arxiv.org/abs/2412.15046  新review
% TODO：还需要把下面的文章都再过一遍，包括他们的引用和谁引用了他们
% TODO: 从 simple-pe 里提取需要引用的文章 [14-30] https://journals.aps.org/prd/pdf/10.1103/PhysRevD.108.082006
% TODO: ReF: https://matheo.uliege.be/bitstream/2268.2/12993/6/report.pdf#page=11.50  % 学习这里的NPE等基础知识
% DONE: 需要提及一下这篇文章： swyft: Truncated marginal neural ratio estimation in python

The detection of gravitational waves from a binary black hole (BBH) merger by the LIGO-Virgo-KAGRA collaboration in 2015 marked a pivotal moment in observational astronomy~\cite{abbott2016observation}. 
Since this groundbreaking discovery, continuous improvements in detector sensitivity and an expanding catalog of detected events~\cite{abbott2019gwtc1,abbott2021gwtc2,abbott2024gwtc21,abbott2023gwtc3} have shifted the focus of astronomy. The field is now increasingly concerned with rapid and detailed parameter estimation as well as comprehensive population-level analyses. These advancements enable astronomers to extract more precise information about the sources of gravitational waves, thereby enhancing our understanding of the universe.

Schutz~\cite{schutz2003Gravitational} argued that gravitational wave detection depends on several critical components: advanced detector technology, precise waveform predictions, robust data analysis methodologies, coincident observations across multiple independent detectors, and complementary observations in electromagnetic astronomy. These elements emphasize the crucial role of data processing in gravitational wave astronomy, which will enable deeper insights into the universe and provide a foundation for addressing the challenges in gravitational wave data analysis.

\subsection{Challenges in Gravitational Wave Data Analysis}

The detection and analysis of gravitational waves fundamentally relies on Bayesian inference to estimate source parameters $\theta$ from detector data $d$. 
According to Bayes’ theorem, the posterior probability distribution is given by
\begin{equation}
p(\theta|d) = \frac{p(d|\theta)p(\theta)}{p(d)} \,,
\end{equation}
where $p(d|\theta)$ is the likelihood, $p(\theta)$ is the prior, and $p(d)$ is the evidence. 
For gravitational wave data analysis, the likelihood typically takes the form
\begin{equation}
p(d|\theta) \propto \exp\left(-\frac{1}{2}\sum_k \frac{|d_k - h_k(\theta)|^2}{S_n(f_k)}\right) \,,
\end{equation}
where $h_k(\theta)$ is the gravitational wave template in the frequency domain and $S_n(f_k)$ is the power spectral density (PSD) of the detector noise.

In this review, we primarily focus on compact binary coalescence sources such as binary neutron star (BNS), BBH, and neutron star–black hole systems—observed in the frequency band of ground-based detectors (10 Hz–1 kHz). In this framework, the following challenges arise:
\begin{itemize}
\item The parameter space $\theta$ has a high number of dimensions (typically 15 parameters or more for binary systems), encompassing mass, spin, orbital parameters, and sky localization. The complexity increases with the inclusion of higher-order modes and environmental effects, making efficient sampling computationally demanding.
\item Generating accurate gravitational wave templates $h_k(\theta)$ is computationally expensive. This challenge is exacerbated when incorporating higher-order modes, precession, eccentricity, or environmental interactions, which require more sophisticated and time-intensive waveform models, such as those derived from numerical relativity.
\item Likelihood evaluations are computationally intensive because of the multiple modes and degenerate parameters in the posterior distribution. These effects complicate the estimation of the likelihood using Monte Carlo sampling methods, which struggle to converge efficiently in such complex landscapes.
\item Non-Gaussian and non-stationary noise transients in the detector data, such as glitches and environmental artifacts, violate the Gaussian noise assumption. These anomalies introduce biases and require advanced noise modeling and mitigation strategies to avoid compromising parameter inference.
\end{itemize}

As reviewed in several studies~\cite{cutler1994gravitational,Christensen2022Parameter,thrane2019introduction}, traditional Markov chain Monte Carlo (MCMC) methods for sampling this posterior typically require hours to days of computation per event.
Because increasing detector sensitivity leads to higher event detection rates, the computational costs for individual event analysis and population studies become prohibitive. Moreover, the complex data processing environment, characterized by high-dimensional parameter spaces and intricate noise characteristics, are highly challenging for Bayesian methods, necessitating the development of more efficient computational techniques and robust data handling strategies.

\subsection{The Rise of Machine Learning in Scientific Computing}

Recent advances in artificial intelligence (AI) and machine learning have significantly reshaped the landscape of scientific computing~\cite{berens2023ai}. In the realm of gravitational wave astronomy, these methods have emerged as formidable tools for tackling a variety of computational challenges~\cite{Cuoco2021enhancing,zhang2024review,cuoco2024applications}. Machine learning techniques, particularly those related to simulation-based inference (SBI), have shown remarkable promise in enhancing gravitational wave parameter estimation, as discussed in detail in Section~\ref{sec:sbi}.

% One pioneering study~\cite{gabbard2022bayesian} employed conditional variational autoencoders (CVAEs) to directly sample posterior distributions, marking a significant advancement in the application of machine learning to gravitational wave analysis. In addition, normalizing flow (NF) models have emerged as powerful tools for parameter estimation, offering robust performance across various scenarios. This approach, initially introduced by Green et al.~\cite{Green2020Gravitational}, provides a compelling alternative to CVAEs. In a subsequent study, Green et al.~\cite{Green2021Complete} applied NF models to the first gravitational wave event, GW150914, sparking a surge of interest and further research in this area. These developments are explored in detail in Section~\ref{sec:app}.

One of the earliest studies to apply SBI methods in gravitational-wave analysis~\cite{Chua2020Learning} developed a variational approach using simple posterior representations, demonstrating the feasibility of learning accurate and even multimodal posteriors from simulations. The method independently derived a loss function structurally equivalent to the neural posterior estimation (NPE) loss~(\ref{eq:npe_loss}) discussed later, and was implemented in the open-source \texttt{truebayes} repository\footnote{\url{https://github.com/vallis/truebayes}}. This work represents a milestone in bridging deep learning with gravitational-wave inference. Around the same time, another pioneering effort~\cite{gabbard2022bayesian} employed conditional variational autoencoders (CVAEs) to directly sample posterior distributions. Subsequently, normalizing flows (NFs) emerged as powerful tools for parameter estimation~\cite{Green2020Gravitational}, offering robust performance and greater flexibility. Their application to GW150914~\cite{Green2021Complete} helped catalyze wider interest in SBI techniques, as explored further in Section~\ref{sec:app}.

This paper presents the first systematic attempt to explore generative models, particularly flow-based deep-learning models, within the context of gravitational wave parameter estimation from an SBI perspective. In Section~\ref{sec:sbi}, we present the technical aspects of SBI and its related architectures. Section ~\ref{sec:app} shifts focus to the scientific applications of these SBI models in various gravitational wave data processing scenarios, providing a comprehensive overview of their impact and potential.

\section{Overview of the SBI methodology} \label{sec:sbi}
%# sbi的介绍, 讲解sbi核心思想以及为什么可以减少计算效率
% The key innovation of SBI is that it does not require an explicit expression for the likelihood function, instead relying solely on the ability to simulate data from the forward model.

% What is SBI
SBI, also known as likelihood-free inference or implicit likelihood inference, has emerged as a powerful paradigm for addressing complex inference problems in gravitational wave astronomy and other fields~\cite{cranmer2020frontier, wang2023sbi, tejero-cantero2020sbi}. Unlike traditional methods, SBI does not require an explicit likelihood expression; instead, it relies on simulators to generate training data. This approach is particularly effective for handling complex, high-dimensional data and models that are challenging to analyze with conventional likelihood-based techniques.

% core idea and advantage of SBI
The core idea of SBI is to create a large training dataset of simulated observations and their corresponding parameters, and then use modern deep-learning techniques to learn the relationships between them. Once trained, these models can rapidly generate posterior samples from new observations, effectively reducing the computational costs of future analyses. A significant advantage of SBI is its ability to learn complex posterior distributions directly from simulated data, which is especially beneficial when dealing with non-stationary or non-Gaussian detector noise, where traditional likelihood evaluations~\cite{PhysRevD.91.042003, Ashton_2019, 10.1093/mnras/staa2850, Speagle_2020} become computationally prohibitive or intractable.

% GW loves the pro of SBI
Gravitational wave data processing greatly benefits from the advantages of SBI. As we enter an era of increased gravitational wave detections, these advances will become particularly crucial. The ability to rapidly generate posterior samples not only accelerates parameter estimation for individual events but also enhances population-level studies through hierarchical inference methods. The computational efficiency of SBI methods is driven by two key factors. First, the costly process of waveform generation and model training is performed only once during the training phase, allowing these expenses to be amortized across future detections. Second, the highly parallelizable nature of neural network computations on modern GPUs results in dramatic speed improvements, often achieving a performance that is thousands of times faster than traditional CPU-based methods. Moreover, while conventional approaches~\cite{Katz2020GPUacceleratedMB, Katz_2022} require additional noise model parameters to be marginalized during inference, SBI methods naturally accommodate such complexities. By incorporating realistic noise conditions into the training data, such as injecting simulated signals into actual instrumental noise, SBI can effectively handle these challenges without additional computational overhead.

\begin{figure*}[htb]
  \centering
  \includegraphics[width=0.9\textwidth]{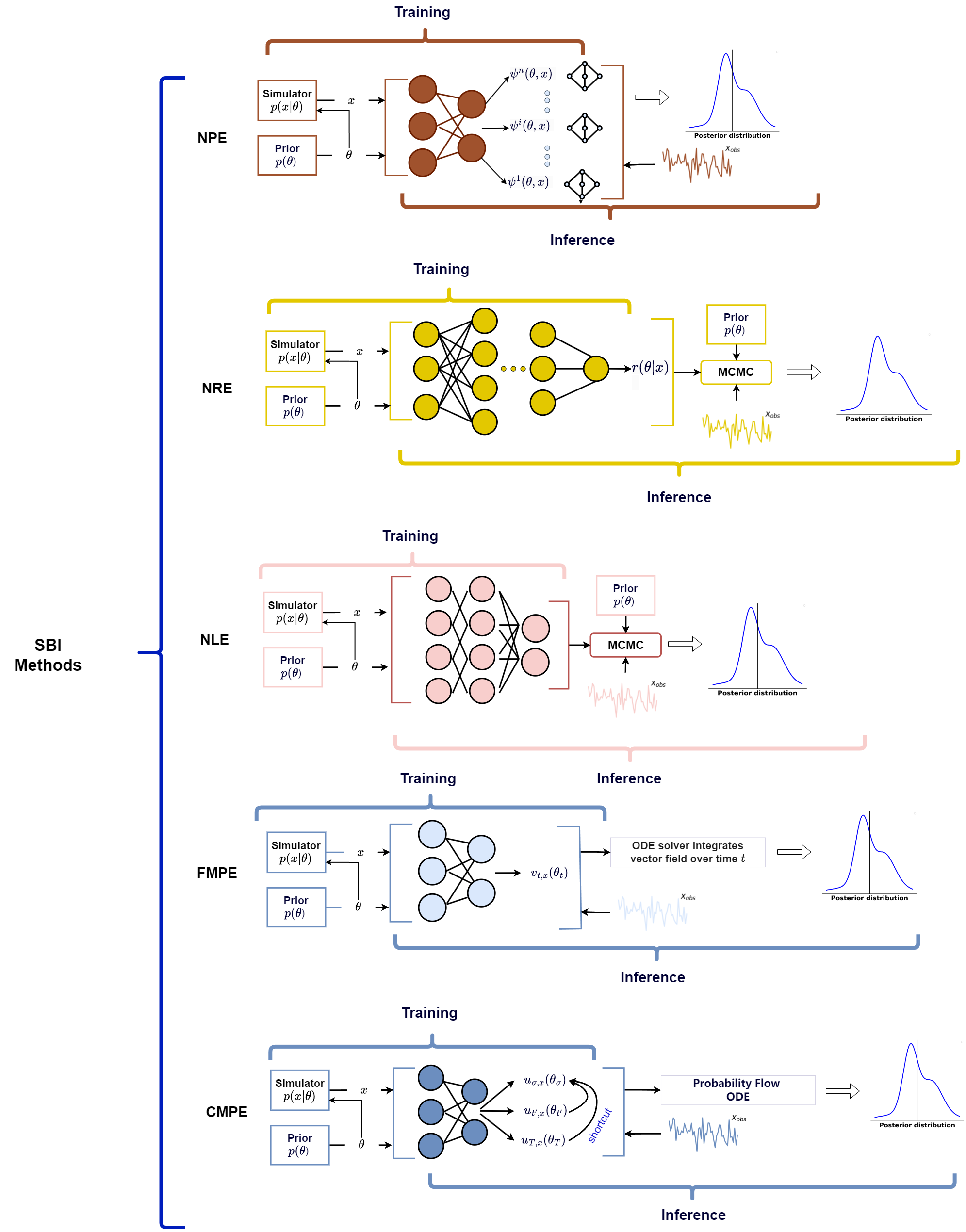}
  \caption{\scriptsize
Overview of five SBI methods—NPE, NRE, NLE, FMPE, and CMPE—designed for efficient Bayesian parameter estimation. Each method includes distinct training and inference stages. NPE trains a neural network to directly approximate the posterior from simulated data. NRE and NLE estimate the likelihood ratio and likelihood function, respectively, and integrate with MCMC for posterior sampling. FMPE  uses an ODE solver guided by a neural network to characterize the parameter posterior. CMPE fits a probability flow with a neural network to sample from posterior distributions. These approaches leverage neural networks to approximate complex posteriors, providing a computationally efficient and flexible alternative to traditional Bayesian inference methods.
  }
  \label{fig:model}
\end{figure*}

The development of SBI has significantly evolved since its early inception as rejection approximate Bayesian computation~\cite{CSILLERY2010410}. Today, SBI encompasses more advanced, neural network-powered, and amortized methods (see Fig.~\ref{fig:model}), such as the following:
\begin{itemize}
    \item \textbf{Neural Posterior Estimation (NPE)}: This method directly learns to generate posterior samples from observed data, enabling rapid inference without explicit likelihood evaluations. It is particularly valuable for real-time analysis and large-scale parameter estimation~\cite{rezende2015variational,papamakarios2016fast,lueckmann2019likelihood,lueckmann2017flexible,greenberg2019automatic}.
     % ~\cite{greenberg2019automatic, deistler2022truncated, pmlr-v202-geffner23a, NEURIPS2023_3663ae53, schmitt2024consistencymodelsscalablefast}.

    \item \textbf{Neural Ratio Estimation (NRE)}: NRE performs inference by estimating the density ratio between the posterior and prior distributions, providing robust performance in scenarios where likelihood ratios are more informative than direct posterior estimation. This method is highly suitable for hypothesis testing and model comparison~\cite{cranmer2015approximating,thomas2022likelihood,hermans2020likelihood,durkan2020contrastive,miller2022contrastive}.
    % ~\cite{hermans2020likelihood, durkan2020contrastive, miller2022contrastive, delaunoy2022towards}.

    \item \textbf{Neural Likelihood Estimation (NLE)}: This approach approximates the likelihood function using neural networks, bridging traditional statistical methods and modern machine learning. It facilitates both parameter inference and uncertainty quantification~\cite{price2018bayesian,papamakarios2019sequential,frazier2023bayesian}.
    
    \item \textbf{Flow Matching Posterior Estimation (FMPE)}: FMPE uses probabilistic streaming models to integrate the joint distributions of parameters and data, supporting precise density estimation, rapid training, and seamless scaling to large architectures. It demonstrates competitive performance on standard SBI benchmarks and exhibits good scalability on challenging scientific problems such as gravitational wave inference~\cite{wildberger2023flow}.

    \item \textbf{Consistency Model Posterior Estimation (CMPE)}: CMPE also employs probabilistic streaming models to distill continuous probability flows, enabling rapid few-shot inference with minimal samples. It is unconstrained and adapts flexibly to the structure of estimation problems, performing well on low-dimensional problems~\cite{schmitt2024consistencymodelsscalablefast}.

\end{itemize}
Recently, there has been substantial interest in applying SBI to high-dimensional parameter spaces. Generative models, such as generative adversarial networks (GANs)~\cite{goodfellow2014generative}, NFs~\cite{dinh2014nice, rezende2015variational, dinh2016density, papamakarios2017masked, huang2018neural, chen2018neural, jaini2019sum, durkan2019neural, wehenkel2019unconstrained, rezende2020normalizing, sick2021deep, meng2020gaussianization, papamakarios2021normalizing}, variational autoencoders~\cite{kingma2013auto,nautiyal2024variational}, Transformers~\cite{gloeckler2024allinone}, continuous normalizing flows (CNFs)~\cite{lipman2023flow, liu2023flow, albergo2023building, tong2023improving, albergo2023stochastic}, score-based and diffusion models~\cite{song2020score, song2020improved,ho2020denoising, song2021maximum, sohl2015deep}, and consistency models (CMs)~\cite{song2023consistency, song2023improved}, offer powerful ways to encode approximate posteriors in these settings.

The following sections explore how these generative models are integrated with inference techniques within the SBI framework. The theoretical foundations and practical implementations of methods such as NPE, NRE, NLE, FMPE, and CMPE are described, highlighting their roles in achieving efficient and accurate parameter estimation.

% Figure \ref{fig:model} describes the five different SBI methods, including NPE, NRE, NLE, FMPE, and CMPE. Each method's training and inference phases are described in detail, highlighting their respective neural network architectures and simulation processes. NPE directly parameterizes the posterior distribution through the neural network, while NRE and NLE perform inference by estimating likelihood ratios and likelihood functions, respectively. FMPE and CMPE use probabilistic streaming models to integrate the joint distributions of the parameters and data to perform posterior inference. These methods provide different strategies for solving complex Bayesian inference problems.

% Figure \ref{fig:ATI} describes five different SBI methods, including NPE, NRE, NLE, FMPE, and CMPE. Each method's training and inference phases are described in detail, highlighting their respective neural network architectures and simulation processes.NPE directly parameterises the posterior distribution through the neural network, while NRE and NLE perform inference by estimating likelihood ratios and likelihood functions, respectively.FMPE and CMPE use probabilistic streaming models to integrate the joint distributions of the parameters and data to perform posterior inference. These methods provide different strategies for solving complex Bayesian inference problems.

\subsection{NPE and NFs}\label{sec:npe}

Within the framework of NFs, a key concept is the transformation of a simple probability distribution $q_0(\theta)$ into a more complex distribution $q(\theta \mid x)$ through an invertible mapping $\psi_x$. This transformation is mathematically described by
\begin{equation}
q(\theta|x) = q_0(\psi_x^{-1}(\theta)) \left| \det \left( \frac{d\psi_x^{-1}}{d\theta} \right) \right| \,.
\end{equation}
Here, $q_0(\theta)$ is the base distribution, $\psi_x$ is the invertible transformation, and $\left|\det\left(\frac{d \psi_x^{-1}}{d \theta}\right)\right|$ is the absolute value of the determinant of the Jacobian matrix of the inverse transformation, which accounts for the change in volume when transforming the probability density.

Building on this theoretical foundation, Papamakarios et al.~\cite{2016arXiv160506376P} introduced NPE as an SBI technique. NPE focuses on training a density estimator $q(\theta \mid x)$ to approximate the true posterior distribution $p(\theta \mid X)$ by minimizing the loss function $L_{\text {NPE }}$, calculated as follows:
\begin{equation}\label{eq:npe_loss}
L_{\text{NPE}} = -\mathbb{E}_{\theta \sim p(\theta)}  \mathbb{E}_{x \sim p(x|\theta)}  \log q(\theta | x) \,,
\end{equation}
where $\mathbb{E}$ denotes the expectation, $\theta$ represents the estimated parameters, and $x$ represents the observed data. In practice, $q(\theta | x)$ is often modeled using NFs to capture complex posterior distributions.

%在NPE的基础上，APT 和 TSNPE 方法被提出，以增强SBI的效率和可扩展性，优化对复杂模型后验分布的估计。
Building on NPE, methods such as automatic posterior transformation~\cite{greenberg2019automatic} and truncated sequential NPE~\cite{deistler2022truncated} have been proposed to enhance the efficiency and scalability of SBI. Automatic posterior transformation automates the transformation of posterior distributions, whereas truncated sequential NPE improves computational efficiency by focusing on relevant parameter spaces. These advancements in NPE and its derivatives demonstrate the power of combining neural networks with probabilistic modeling to tackle complex Bayesian inference problems effectively.

% Neural Posterior Estimation (NPE) represents a transformative approach in gravitational wave data analysis, enabling direct learning of posterior distributions from observed data. The groundbreaking work by~\cite{gabbard2022bayesian} in Nature marked a watershed moment, demonstrating that NPE could achieve accuracy comparable to traditional MCMC methods while reducing inference time by several orders of magnitude. This achievement opened new possibilities for real-time gravitational wave analysis and rapid electromagnetic follow-up observations.
A significant advancement in the development of NPE has been the deep inference for gravitational-wave observations (DINGO) framework~\cite{Green2021Complete,Dax2021Real,Dax2023Neural,Green2020Gravitational,Wildberger2023Adapting,Dax2023Neural,Leyde2024Gravitational,gupte2024evidence,dax2024real}. This framework leverages the flexibility and power of NFs for probabilistic modeling, demonstrating broad applicability beyond NPE. Some specific areas where NFs have shown promise are as follows:

\begin{itemize}
\item \textbf{Embedding-based Approaches}: These methods utilize embeddings to enhance statistical inference and premerger analysis, providing rapid insights into gravitational wave data processing~\cite{Shen2022Statistically,Chatterjee2023Premerger,Chatterjee2023Rapid}.
\item \textbf{Applications to Specific Source Types}:
\begin{itemize}
\item \textbf{Binary Close Encounters}: NF has been applied to model the dynamics of binary systems during close encounters, improving inference accuracy~\cite{DeSanti2024deep}.
\item \textbf{Glitch-Robust Inference}: Techniques have been developed to enhance robustness against glitches in gravitational wave data, ensuring reliable parameter estimation~\cite{xiong2024robust,Sun_2024}.
\item \textbf{Intermediate-Mass Mergers}: NF has facilitated the simulation and analysis of intermediate-mass black hole mergers, a challenging area in gravitational wave astronomy~\cite{raymond2024simulation}.
\end{itemize}
\item \textbf{Enhanced Sampling Methods}:
\begin{itemize}
\item \textbf{Flow-Enhanced MCMC}: This approach integrates NF with Markov Chain Monte Carlo (MCMC) methods to accelerate convergence and improve sampling efficiency~\cite{Wong2023Fast,Wouters2024robust}.
\item \textbf{Flow-Enhanced Nested Sampling}: NF has been used to enhance nested sampling techniques, providing faster and more robust posterior inference~\cite{Williams2021Nested,Roulet2024fast,lange2023NAUTILUS}.
\item \textbf{Improvements to Hamiltonian Monte Carlo Methods}: These methods have been improved using NFs, leading to more efficient exploration of parameter spaces~\cite{edwards2023ripple}.
\end{itemize}
\end{itemize}
These applications highlight the versatility and effectiveness of NPE and NFs in advancing gravitational wave data analysis and inference.

\subsection{NRE}

NRE is a machine learning approach that trains a classifier network to distinguish between sample pairs $(\theta, X)$ drawn from the joint distribution $p(\theta, x)$ and those from the product of the marginal distributions $p(\theta) p(x)$. 
The core idea is to estimate the likelihood ratio, which is the ratio of the posterior probability of the parameters $\boldsymbol{\theta}$ given the observed data $X$ to their prior probability.

The objective of NRE is to train a network $d_\phi(\theta, x)$ that emulates the decision function of a Bayesian optimal classifier, defined by the following likelihood-to-evidence ratio:
\begin{equation}
r(\theta, x) = \frac{p(\theta | x)}{p(\theta)} = \frac{p(x | \theta)}{p(x)} \,.
\end{equation}
Here, $p(x | \theta)$ is the conditional probability density of the observed data $X$ given parameters $\theta$, and $p(X)$ is the marginal probability density of the observed data.

Training NRE involves minimizing the following loss function:
\begin{equation}
\arg\min_\phi \frac{1}{2} \mathbb{E}_{p(\theta, x)} \left[ -\log q \right] + \frac{1}{2} \mathbb{E}_{p(\theta)p(x)} \left[ -\log (1 - q) \right] \,,
\end{equation}
where
\begin{equation}
q = d_\phi(\theta, x).
\end{equation}
The loss function consists of two terms: the negative log-likelihood\footnote{$-\log p$ is the negative log-likelihood function, which measures the discrepancy between the predicted probability output by the network $p$ and the actual probability.} for joint distribution $p(\theta, x)$ and the negative log-likelihood for the product of the marginals $p(\theta) p(x)$. The goal is to train the network $d_\phi(\theta, x)$ to output higher probabilities for samples from the joint distribution and lower probabilities for samples from the product of the marginals, which is effectively learning to estimate the likelihood ratio.

The decision function of the Bayesian optimal classifier is given by
\begin{equation}
d(\theta, x) = \frac{p(\theta, x)}{p(\theta, x) + p(\theta)p(x)} \,.
\end{equation}
This function defines the likelihood-to-evidence ratio, expressed as:
\begin{equation}
r(\theta, x) = \frac{d(\theta, x)}{1 - d(\theta, x)} = \frac{p(\theta, x)}{p(\theta)p(x)} \,.
\end{equation}
Here, $d(\theta, x)$ represents the decision function indicating the probability that a sample pair $(\theta, x)$ is drawn from the joint distribution $p(\theta, x)$ rather than the product of the marginals $p(\theta)p(x)$. The likelihood-to-evidence ratio $r(\theta, x)$ is the ratio of the joint probability to the product of the marginal probabilities, effectively capturing the relative evidence provided by the data $x$ for the parameters $\theta$.

To ensure numerical stability, especially when $d_\phi(\theta, x) \to 0$, NRE networks return the logits of the class predictions, which are the logarithms of the odds; that is,
\begin{equation}
\text{logit}(d_\phi(\theta, x)) = \log r_\phi(\theta, x)\,.
\end{equation}
Here $r_\phi(\theta, x)$ is the likelihood ratio estimated by the network. Taking the logarithm provides a numerically stable representation of the odds, which is useful for handling extreme values of $r_\phi(\theta, x)$. The logit function transforms probability estimates into a more manageable form for learning and computation.

% Below is a comprehensive table~\ref{table:nre} listing several important relevant NRE variants a and their corresponding academic papers. 
These models have demonstrated exceptional performance across various application scenarios, significantly enhancing the efficiency and accuracy of SBI. Among them, truncated marginal NRE (TMNRE)~\cite{miller2021truncated}, also known as Swyft, has emerged as a leading method and has been widely applied across multiple scientific fields~\cite{anau2023estimating, karchev2023sicret, coogan2022one, montel2022detection, gagnon2023debiasing, bhardwaj2023sequential, alvey2023albatross}. The renowned Peregrine project~\cite{bhardwaj2023sequential}, which focuses on sequential SBI for gravitational wave signals, has successfully adopted the TMNRE method.
TMNRE's ability to efficiently handle complex parameter spaces and its robustness in various inference tasks make it particularly suitable for gravitational wave data analysis. By leveraging the strengths of TMNRE, researchers can achieve significant reductions in computational costs while maintaining high accuracy in parameter estimation, making it a valuable tool in the field of gravitational wave astronomy.

\subsection{NLE}
In Bayesian inference, when the likelihood function $p(x|\theta)$ is challenging to compute directly, traditional inference methods often become impractical. To address this issue, Papamakarios et al.~\cite{papamakarios2019sequential} introduced Sequential Neural Likelihood (SNL), a novel approach that focuses on directly modeling the likelihood function rather than the posterior distribution, thereby avoiding biases introduced by the proposal distribution.

The SNL method begins by defining a proposal distribution $p_{\sim}(\theta)$, which is used to generate parameter samples $\theta_n$. Subsequently, data $x_n$ are simulated based on these parameter samples. These sample pairs $\{\theta_n, x_n\}_{n=1}^N$ form
the joint distribution $p_{\sim}(\theta, x) = p(x|\theta)p_{\sim}(\theta)$. Given this foundation, SNL trains a conditional neural density estimator $q_{\phi}(x|\theta)$, which models the conditional probability density of data $x$ given parameters $\theta$.

% \(\sum_{n} \log q_{\phi}(x_n|\theta_n)\)

For a sufficiently large sample size $N$, maximizing the total log-likelihood is equivalent to maximizing the expected log-likelihood $\mathbb{E}_{p_{\sim}(\theta, x)}[\log q_{\phi}(x|\theta)]$. This expectation can be expressed as
\begin{equation}
\mathbb{E}_{p_{\sim}(\theta, x)}[\log q_{\phi}(x|\theta)] = -\mathbb{E}_{p_{\sim}(\theta)}[K] + \text{const} \,,
\end{equation}
where
\begin{equation}\label{eq:K}
    K = D_{KL}(p(x|\theta) \| q_{\phi}(x|\theta)) \,,
\end{equation}
In Eq. (\ref{eq:K}),  $D_{KL}(\cdot \| \cdot)$ is the Kullback–Leibler (KL) divergence. The above quantity is maximized when the KL divergence over the support of $p_{\sim}(\theta)$ is zero, i.e., when $q_{\phi}(x|\theta) = p(x|\theta)$ for all $\theta$ where $p_{\sim}(\theta) > 0$. 

By directly learning the likelihood function, SNL provides a flexible and powerful framework for Bayesian inference, particularly in scenarios where traditional methods struggle because of complex or intractable likelihoods. This approach allows for more accurate parameter estimation and uncertainty quantification, making it a valuable tool in the analysis of complex gravitational wave signals.

\subsection{Flow Matching (FM) and CNFs}

CNFs facilitate the transformation of a base distribution into a more complex distribution by mapping the transformation across a continuous time interval $t \in[0,1]$. 
At each time $t$, the flow is defined by a vector field $v_{t, x}$ that indicates the velocity at which samples move along their trajectories. This vector field is governed by the following ordinary differential equation (ODE):
\begin{equation}
\frac{d}{dt} \psi_{t,x}(\theta) = v_{t,x}(\psi_{t,x}(\theta)) \,,
\end{equation}
where $\psi_{t, x}(\theta)$describes the transition from the base distribution to the target distribution. The shift from the base distribution (at $t=0$ ) to the posterior distribution (at $t=1$ ) is achieved using integration. Training CNFs typically starts with a base distribution and uses an ODE solver to obtain the posterior distribution, optimizing it to match the empirical data distribution by minimizing a divergence metric such as the KL divergence.

However, because of the complexity of many intermediate paths, whether by sampling or likelihood calculation, multiple runs of the ODE are required, which can be computationally demanding. The concept of flow matching (FM) provides a training objective for CNFs that Dax et al.~\cite{dax2023flowmatchingscalablesimulationbased} have adapted for SBI. The loss function for FMPE is formulated as
\begin{equation}
L_{FMPE} = \mathbb{E}_{t \sim p(t), \theta_1 \sim p(\theta), x \sim p(x|\theta_1), \theta_t \sim p_t(\theta_t|\theta_1)}  \parallel r \parallel^2 \,,
\end{equation}
Where $r =  v_{t,x}(\theta_t) - u_t(\theta_t|\theta_1)$ and $t \sim p(t)$ represents sampling from time $t = 0$ to $t = 1$, usually uniformly distributed.
$\theta \sim p(\theta)$ represents the true posterior distribution for sampling $\theta$.
Data $x \sim p(x|\theta_1)$  represents the simulated data corresponding to $\theta$. 
Distribution $\theta_t \sim p_t(\theta_t|\theta_1)$ represents the $\theta_t$ constructed by stochastic interpolation.
In this context, \( v_{t,x}(\theta_t) \) is the vector field that directs the path to the desired probability distribution, and \( u_t(\theta_t|\theta_1) \) is the vector field conditioned on the sample. 
FMPE defines the $u_t(\theta_t|\theta_1)$ vector field as
\begin{equation}
u_t(\theta|\theta_1) = \frac{\theta_1 - (1-\sigma)\theta_{0}}{1 - (1 - \sigma)t} \,.
\end{equation}
This formulation allows FMPE to efficiently guide the transformation of distributions, providing a robust framework for Bayesian inference in complex models.

Recent work~\cite{wildberger2023flow, Liang2024Rapid} has demonstrated that FM can achieve state-of-the-art performance with significantly fewer function evaluations. This efficiency makes FM particularly promising for real-time gravitational wave inference, where rapid posterior estimation is crucial for multi-messenger follow-up observations. The method's ability to capture complex topological features of the posterior distribution also makes it well-suited for analyzing signals from exotic sources and testing general relativity.

The current application of FMPE in the field of gravitational waves primarily builds on the method proposed by Lipman et al.~\cite{lipman2023flow}. Given the rapid advancements in FM research within the imaging field, exploring the extension of these innovative methods to SBI in gravitational wave astronomy holds significant potential. Such exploration could lead to enhanced techniques for parameter estimation and signal analysis, further advancing our understanding of gravitational wave phenomena and their implications for fundamental physics.

\subsection{CMPE}

To address the large amount of time consumed in the inference process of diffusion models, Song et al. proposed the CM, which is a new type of generative model. These models effectively overcome the limitations of traditional diffusion models in terms of sampling speed by supporting both single-step and multi-step sampling. The training objective of the CM is to learn a consistency function $f_{\phi}(\theta_t,t,x) \rightarrow \theta_{\sigma}$ that maps points $\{\theta_{t}\}_{t\in[T,\sigma]}$ in the probability flow directly to the origin $\theta_{\sigma}$ of the trajectory as
\begin{equation}
       f_{\phi}(\theta_t,t,x)  = c_{skip}(t)\theta + c_{out}(t)F_{\phi}(\theta,t,x) \,,
\end{equation}
where  $c_{skip}(t)$ and $c_{out}(t)$ are differentiable functions satisfying $c_{skip}(\sigma) = 1$ and $c_{out}(\sigma) = 0$. Function $F_{\phi}(\theta,t,x)$ builds a score-based diffusion model structure to achieve this mapping.

Schmitt et al. extended the unconditional training objective from Song by introducing the conditioning variable x to cater to the SBI setting~\cite{schmitt2024consistencymodelsscalablefast} as follows:
\begin{equation}
    L_{CMPE}(\phi, \phi^-) = \mathbb{E} \left[ \lambda(t_i)k \right] \ \,,
\end{equation}
where $k = d(u(\phi, t_{i+1}, x), u(\phi^-, t_i, x))$, $\lambda(t)$ is a weighting function, and $d(u, v)$ is a distance metric.
The function $u(\phi, t; x)$ is defined as:
\begin{equation}
    u(\phi, t; x) = f_{\phi}(\theta + tz, t; x) \,,
\end{equation}
where $z$ is Gaussian noise, and $\phi^{-}$ is a copy of the student parameter, held constant at each step by the stop-gradient operator.

This approach allows CMPE to efficiently handle complex inference tasks by leveraging the strengths of CMs, providing a robust framework for Bayesian parameter estimation in challenging scenarios.

Note that CM techniques have not yet been applied to the field of gravitational wave analysis. However, given the burgeoning research and advancements in CM methodologies\cite{Song2023ConsistencyM, song2023improved,Lu2024SimplifyingSA, Kim2023ConsistencyTM, Geng2024ConsistencyMM}, the exploration of their potential application in gravitational wave data processing holds significant promise. The integration of CM techniques could offer novel insights and improvements in the efficiency and accuracy of gravitational wave parameter estimation, making it a compelling area for future research.

\begin{table*}[h!]
\centering
\scriptsize
\begin{tabularx}{\textwidth}{@{}>{\centering\arraybackslash}m{0.15\textwidth} >{\centering\arraybackslash}m{0.3\textwidth} >{\centering\arraybackslash}m{0.45\textwidth} >{\centering\arraybackslash}m{0.1\textwidth}@{}}
\toprule
\textbf{Research Field} & \textbf{Challenges} & \textbf{Scientific Significance} & \textbf{Related Work} \\ \midrule
\multirow{3}{0.15\textwidth}    {\centering  \\[10ex]  \textbf{Single-Source Parameter Estimation and Binary Systems}}  & 
\textbf{BBH Systems:} 
BBH parameter estimation involves
15+ dimensions (masses, spins, orbital parameters,
etc.). Traditional Bayesian methods (e.g., MCMC)
are computationally expensive, limiting real-time
analysis.

 & 
Rapid sky localization (e.g., right ascension, distance) guides real-time follow-up observations with electromagnetic/neutrino telescopes, probing the electromagnetic counterparts and host galaxies of BBH mergers. Rapid inference is essential for probing high-redshift BBH populations and testing gravitational-wave cosmology. & 
\cite{Green2020Gravitational}, \cite{Green2021Complete}, \cite{Dax2021Real}, \cite{Dax2023Neural}, \cite{wildberger2023flow}, \cite{Chatterjee2023Rapid}, \cite{raymond2024simulation}, \cite{xiong2024robust},\cite{Sun_2024}, \cite{DeSanti2024deep}, \cite{Wong2023Fast}, \cite{Williams_2023}, \cite{Kuo2022Conditional}. \\ \cmidrule{2-4} 
 & \textbf{BNS Systems:} Approximate low-latency methods (e.g., BAYESTAR) discard phase information, increasing localization errors by 30\%. Long signals exacerbate non-Gaussian noise effects, raising systematic error risks. & 
Enables full-parameter inference within one second, critical for rapid electromagnetic follow-up. Provides alerts minutes before merger, enabling searches for precursor electromagnetic signals. Enables real-time EOS likelihood evaluation via parameter conditioning, advancing nuclear physics studies.& 
\cite{dax2024real}, \cite{van2024pre}, \cite{Wouters2024robust}, \cite{mcginn2024rapid}. \\ \cmidrule{2-4} 
 & \textbf{Forecasting Future PSDs:} 
PSD non-stationarity complicates real-time modeling; conventional
methods sacrifice accuracy for speed in low-latency scenarios. & 
Enables reliable inference under dynamic noise, critical for real-time multi-messenger observations of extreme events (e.g., neutron star mergers). Provides a core methodology for the data processing of the high-event-rate of third-generation detectors, advancing gravitational-wave cosmology.& 
\cite{Wildberger2023Adapting}. \\ \midrule

\multirow{2}{0.15\textwidth}{ \centering \\[5ex] \textbf{Overlapping Gravitational Wave Signals}} & 
\scriptsize \textbf{Ground-based Gravitational Wave Signals:}
Overlapping short-duration merger signals (e.g., compact binary coalescences) distort waveform features, causing systematic shifts in the joint posterior distributions of key parameters (masses, spins), especially in low signal-to-noise scenarios.& 
\scriptsize  Millisecond-scale overlap resolution allows synchronized electromagnetic/neutrino telescope follow-up, capturing kilonovae and gamma-ray bursts associated with compact binary mergers. Precise signal separation reveals faint binary populations inaccessible to traditional methods, constraining stellar evolution and gravitational wave cosmology models. & 
\scriptsize \cite{Langendorff2023Normalizing}, \cite{alvey2023things}. \\ \cmidrule{2-4} 
 & \scriptsize \textbf{Space-based Gravitational Wave Signals:} Persistent overlap of millions of gravitational wave signals (e.g., galactic binary confusion noise) in data streams leads to exponential computational complexity for traditional template-based methods.& 
\scriptsize Detection of supermassive black hole mergers provides direct insights into the dynamics of the Epoch of Reionization, complementing optical surveys. Precision waveform phase measurements from EMRIs validate general relativity in strong-field regimes or reveal new physics.& 
\scriptsize \cite{Ruan2023Parameter}, \cite{du2024advancing}, \cite{Liang2024Rapid}. \\ \midrule

\multirow{2}{0.15\textwidth}{ \centering  \\[2ex] \textbf{Testing General Relativity and Cosmology}} & 
\scriptsize \textbf{Tests of General Relativity:} 
Current template banks (e.g., SEOBNR, IMRPPhenom) are general relativity-based, lacking parametric coverage for alternative theories, potentially misclassifying beyond-general relativity signals as noise anomalies.& 
\scriptsize Detecting deviations from general relativity predictions (e.g., frequency-dependent propagation speed, extra polarization modes) could falsify existing frameworks. & 
\scriptsize \cite{Crisostomi2023Neural}, \cite{Pacilio2024Simulation}. \\ \cmidrule{2-4} 
 & \scriptsize \textbf{Testing Cosmology:}
Traditional methods (e.g., ladder likelihood) face prohibitive computational costs in high-dimensional cosmological parameter spaces (e.g., $H_{0}$, dark energy equation of state $w$) and struggle with parameter degeneracies. & 
\scriptsize Independent $H_{0}$ measurements via gravitational wave standard sirens bypass systematic errors in cosmic distance ladders, offering new insights into the Hubble tension.
Normalizing flow-based generative models efficiently model non-Gaussian cosmological fields, improving statistical inference on key probes like halo mass functions and baryon acoustic oscillations. & 
\scriptsize \cite{Stachurski2024Cosmological}, \cite{rouhiainen2021normalizingflowsrandomfields}. \\ \midrule

\multirow{2}{0.15\textwidth}{ \centering \\[5ex]  \textbf{Population Studies}} & 
\scriptsize \textbf{Gravitational Wave Population Inference:}
High-dimensional parameter spaces and computational complexity mean that the analysis of thousands of events with correlated parameters (masses, spins, redshifts) demands scalable algorithms for high-dimensional data.& 
\scriptsize Probing compact binary formation reveals formation channels (stellar evolution, dynamical capture) and advances the understanding of astrophysical processes. & 
\scriptsize \cite{Wong_2020}, \cite{Wong2020GravitationalwavePI}, \cite{PhysRevD.102.023025}, \cite{Mould_2022}, \cite{ruhe2022normalizingflowshierarchicalbayesian}, \cite{Ray_2023}, \cite{Leyde_2024}. \\ \cmidrule{2-4} 
 & \scriptsize \textbf{Gravitational Wave Population Discovery and Robustness Testing:}
Ensuring statistical methods remain reliable under non-Gaussian noise, calibration errors, or incomplete astrophysical models is critical for avoiding false discoveries. & 
\scriptsize
Population discovery enables tests of exotic formation channels (e.g., primordial black holes, dynamical capture) and probes physics beyond standard stellar evolution. Rigorous robustness testing ensures methods scale to next-generation observatories (e.g., Einstein Telescope, Cosmic Explorer) with higher sensitivity and increased event rates. & 
\scriptsize \cite{Tiwari_2021}, \cite{Cheung2021TestingTR}, \cite{wong2022automateddiscoveryinterpretablegravitationalwave}, \cite{PhysRevD.100.083015}. \\ \bottomrule
\end{tabularx}
\caption{\scriptsize Summary of key research areas in gravitational-wave astronomy where AI methods are applied, highlighting the challenges, scientific significance, and related work across single-source parameter estimation (BBH/BNS systems), overlapping signal analysis (ground/space-based detectors), fundamental theory testing (general relativity/cosmology), and population studies.}
\label{table:ATI}
\end{table*}

\subsection{Methodological Limitations}
Existing validation protocols predominantly rely on localized comparisons with traditional MCMC results\cite{doi:10.1073/pnas.1912789117}, and they lack systematic evaluation metrics for high-dimensional parameter spaces. Neural networks may exhibit progressive biases in extreme parameter regions that remain undetected under current verification paradigms~\cite{NEURIPS2022_9278abf0}. 
While achieving accelerations of $10^2 - 10^3$ in per-event inference\cite{Dax2021Real,dax2024real,dax2023flowmatchingscalablesimulationbased,Dax2023Neural}, the total computational cost becomes comparable to that of traditional methods when incorporating waveform model updates (e.g., higher-order modes~\cite{PhysRevD.110.044063, PhysRevD.110.084035, Wadekar:2023gea} ). For third-generation detectors, which are expected to observe $10^3$ annual events, repetitive retraining requirements will impose prohibitive resource demands~\cite{Green2021Complete}. 

Training data generated under idealized simulations neglect environmental interactions and exotic compact object properties. This induces unexplained biases when analyzing real data containing anomalous features~\cite{Liang2024Rapid}. More insidiously, neural networks exhibit intrinsic inductive biases that produce unphysical predictions in sparse data regions despite sufficient training samples~\cite{Liang2024Rapid}.
Real-time deployment requires integration with preprocessing pipelines (data cleaning, PSD estimation), where cumulative latency often exceeds theoretical inference time~\cite{Dax2021Real,Liang2024Rapid,dax2024real}.
%------------------------------------------------------------------------------
\section{Applications and Implications}\label{sec:app}
%------------------------------------------------------------------------------
% 从 simple-pe 里提取需要引用的文章 [14-30] https://journals.aps.org/prd/pdf/10.1103/PhysRevD.108.082006

In the field of gravitational wave data analysis, SBI methods have received widespread attention for their efficiency and flexibility in dealing with complex Bayesian inference problems. However, SBI methods also face a series of challenges in practical applications, such as how to achieve fast and accurate parameter estimation and how to handle overlapping gravitational wave signals. To overcome these challenges, researchers have proposed a variety of innovative solutions and have achieved significant progress in improving the efficiency and accuracy of inference. Table \ref{table:ATI} summarizes the main challenges faced by current machine learning methods used in gravitational wave data analysis and the corresponding research progress, providing an overview of the latest trends and technological developments in this field.

% \begin{figure*}[ht!]
%   \centering
%   \includegraphics[width=0.9\textwidth]{ATI.png}
%   % \captionsetup{font={footnotesize,stretch=0.4}}
%   \caption{\scriptsize The figure outlines the machine learning methods applied in the estimation of gravitational wave parameters, which fall into two main categories: the SBI methods and the other methods. The figure lists the specific research papers that have been used in the context of a particular challenge and the methods they have employed.  Next to each method, the techniques they use are labelled, e.g., NF, MCMC, IS, etc.
%   }
%   \label{fig:ATI}
% \end{figure*}

\subsection{Single-Source Parameter Estimation and Binary Systems}

Early work in gravitational wave data analysis using deep learning began around 2018, when pioneering studies employed convolutional neural networks and formulated signal detection as a binary classification task~\cite{george2018deep,gabbard2018matching}. Researchers quickly recognized the immense promise of deep-learning techniques and began exploring their application to parameter estimation—the ``holy grail" of gravitational wave data analysis.

In the early stages of research, parameter inference was typically performed via point estimation methods using convolutional neural network-based architectures~\cite{Kolmus2022Fast, beveridge2023detection, Sasaoka2022Localization, Nunes2024Deep, AndresCarcasona2023Fast, mcleod2022rapid}. However, since scientific discoveries require rigorous uncertainty quantification methods that yield full posterior distributions instead of single-point estimates, the focus soon shifted toward Bayesian interval estimation.

A seminal contribution in this area was provided by Gabbard et al.~\cite{gabbard2022bayesian}, who implemented a conditional variational autoencoder framework to infer the complete 15-dimensional posterior probability density for BBH systems. This approach achieved inference speeds that were nearly 4–5 orders of magnitude faster than traditional sampling methods (such as MCMC or nested sampling), while producing posterior distributions that closely matched conventional results.

Shortly after, Green et al.~\cite{Green2020Gravitational} compared conditional variational autoencoder-based methods with those based on NFs for BBH parameter estimation. They found that flow-based models offer more robust performance, particularly in generating smoother posterior distributions. Building on this, Green et al.~\cite{Green2021Complete} demonstrated the efficacy of NF models in realistic noise environments. In their analysis of the first detected event, GW150914, they showed that a flow-based model could produce tens of thousands of posterior samples in just a few seconds, with results that had excellent agreement with MCMC estimates. Their early public release\footnote{\url{https://github.com/stephengreen/lfi-gw/}} has become a primary resource, enabling further testing and adaptation of the method to other BBH events~\cite{wang2021sampling}.

\begin{figure}[ht!]
  \centering
  \includegraphics[width=0.5\textwidth]{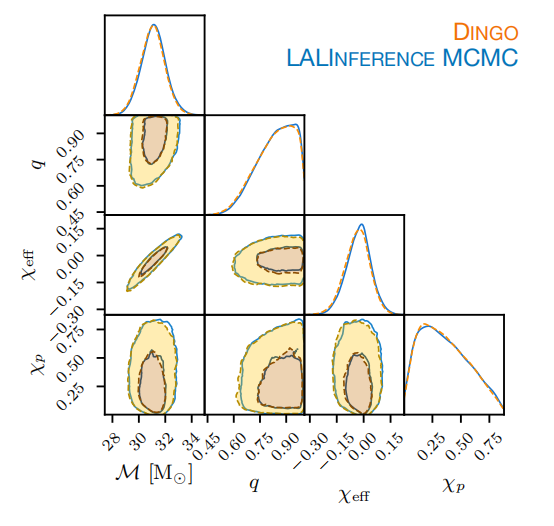}
  \caption{\scriptsize
  Comparison of the marginalized posterior distributions for event GW150914 obtained using DINGO (orange) versus LALInference MCMC (blue), underscoring the efficiency and accuracy of the DINGO framework in real-time gravitational wave parameter estimation. (Adapted from Ref.~\cite{Dax2021Real})
  }
  \label{fig:dingo}
\end{figure}

One limitation of flow-based models is the requirement that the PSD must be sampled near the target event during training, which means that their general applicability is more restricted than that of MCMC. To mitigate this limitation, the DINGO framework~\footnote{\url{https://github.com/dingo-gw/dingo}} was developed by Dax et al.~\cite{Dax2021Real}. DINGO leverages PSD samples drawn from the full span of observational data (e.g., during O1/O2) to train the model. This approach enables the precise characterization of the posterior distributions for all BBH events, as illustrated in Fig.~\ref{fig:dingo}. Furthermore, subsequent studies~\cite{Wildberger2023Adapting} have introduced methods to adapt to noise distribution shifts without the need for retraining, thereby positioning DINGO as a promising tool for online, real-time statistical inference in future ground-based gravitational wave detectors.

A subsequent milestone from the same team is DINGO-IS~\cite{Dax2023Neural}, which refines parameter estimation by reweighting samples obtained via NPE (as reviewed in Section~\ref{sec:npe}). This reweighting not only verifies and corrects the machine-learned posterior but also yields an unbiased estimate of the Bayesian evidence because of the normalization of the proposal distribution.

\begin{figure}[ht!]
  \centering
  \includegraphics[width=0.5\textwidth]{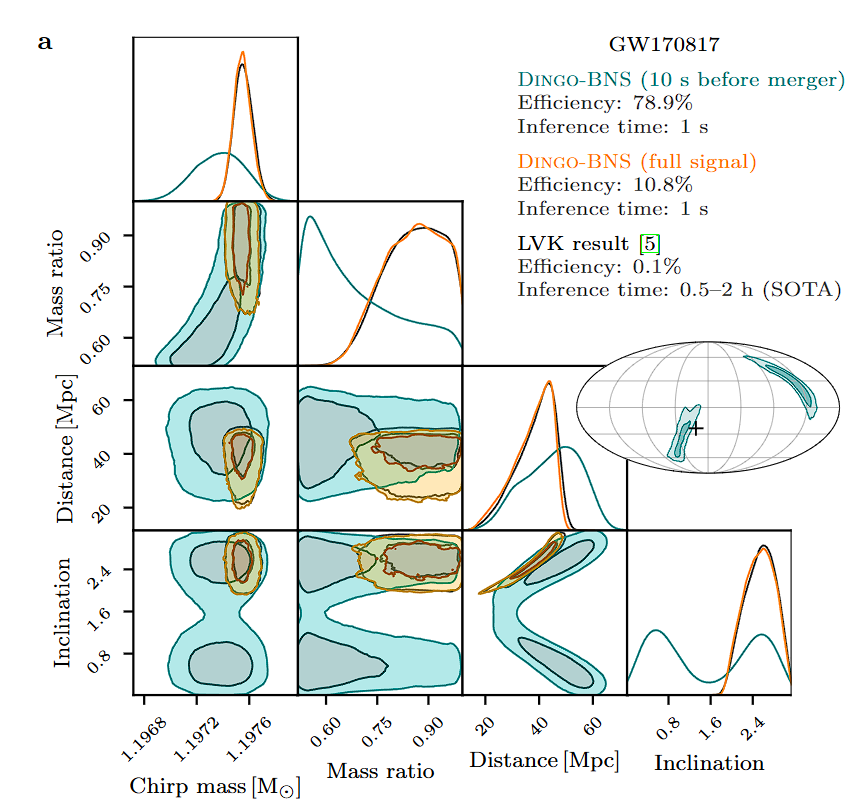}
  \caption{\scriptsize
  The DINGO-BNS algorithm can estimate the parameters for BNS systems within one second (orange), yielding results consistent with LIGO-Virgo-KAGRA analyses (black) while operating three orders of magnitude faster than traditional methods. (Reproduced from Ref.~\cite{dax2024real})
  }
  \label{fig:dingo-bns}
\end{figure}

More recently, research has extended the DINGO framework to BNS systems. Dax et al.~\cite{dax2024real} introduced DINGO-BNS, a method tailored for BNS events that features a real-time inference pipeline capable of generating the complete posterior distribution within seconds of detection, as illustrated in Fig.~\ref{fig:dingo-bns}. In addition, DINGO has been successfully applied to population-level inference of source distributions~\cite{Leyde2024Gravitational} and parameter estimation for eccentric binary systems~\cite{gupte2024evidence}.

% Parallel work has also addressed the unique challenges posed by BNS systems, which involve complex matter effects such as tidal deformability. Wouters et al.~\cite{Wouters2024robust} developed a method capable of robust parameter estimation for BNS inspirals within minutes, incorporating both amplitude and phase corrections from tidal effects. Their approach demonstrated an exceptional ability to accurately estimate the tidal deformability parameter \(\tilde{\Lambda}\) -- a crucial parameter for constraining the neutron star equation of state. Building on this, van et al.~\cite{van2024pre} showcased early warning capabilities by achieving pre-merger detection and characterization of inspiraling BNS systems using NPE, thereby enabling timely electromagnetic follow-up observations.

Parallel work has also addressed the unique challenges posed by BNS systems, which involve complex matter effects such as tidal deformability. Wouters et al.~\cite{Wouters2024robust} developed a method capable of robust parameter estimation for BNS inspirals within minutes, incorporating both amplitude and phase corrections from tidal effects. Their approach demonstrated an exceptional ability to accurately estimate the tidal deformability \(\tilde{\Lambda}\) -- a crucial parameter for constraining the neutron star equation of state. Building on this, van Straalen et al.~\cite{van2024pre} demonstrated early warning capabilities by achieving the pre-merger detection and characterization of inspiraling BNS systems using NPE, thereby enabling timely electromagnetic follow-up observations. Most recently, McGinn et al.~\cite{mcginn2024rapid} introduced ASTREOS, a rapid approach based on NFs for inferring the neutron star equation of state. Their method produces results consistent with the LIGO-Virgo collaboration findings, but it requires less than one second to generate confidence intervals, representing a significant advancement in both speed and flexibility for neutron star physics studies.

% %%

% todo:Transformer + BNS

\subsection{Overlapping Signals}

The analysis of overlapping gravitational wave signals presents unique challenges because of the increased complexity involved in disentangling multiple sources~\cite{niu2024extracting}. As detector sensitivity improves—especially with the advent of third-generation detectors—the likelihood of observing overlapping signals increases significantly. In space-based gravitational wave detection, this issue becomes even more severe~\cite{Bayle2022Overview,perir2022roadmap}: data streams may contain on the order of \(10^5\) to \(10^6\) overlapping signals from different sources, posing unprecedented challenges for data processing and analysis.

\begin{figure}[ht!]
  \centering
  \includegraphics[width=0.5\textwidth]{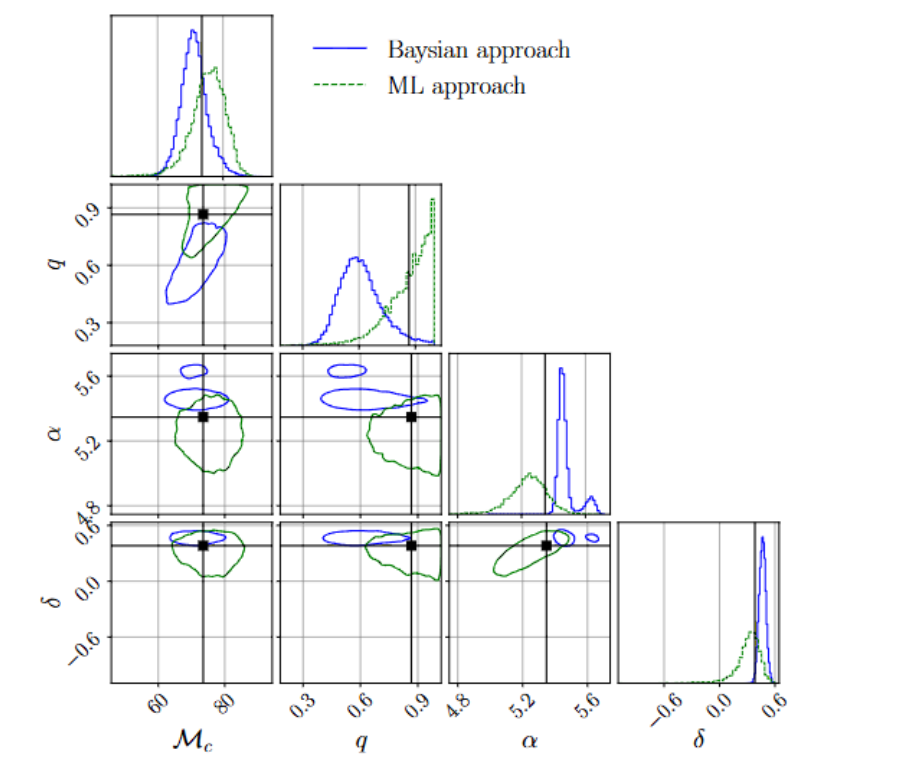}
  \caption{\scriptsize
  Comparison of posterior distributions for overlapping gravitational wave signals obtained by the approach of Langendorff et al.~\cite{Langendorff2023Normalizing} and the method presented in Ref.~\cite{Janquart_2023}. Langendorff et al.’s posteriors are generally broader but consistently encapsulate the injected values within the 90\% confidence interval, indicating the potential for further refinement via importance sampling. (Image from Ref.~\cite{Langendorff2023Normalizing})
  }
  \label{fig:overlap}
\end{figure}

Early work by Langendorff et al.~\cite{Langendorff2023Normalizing} demonstrated that CNFs can address the challenges of parameter inference for overlapping gravitational wave signals in third-generation detectors. Their study emphasized a critical point: traditional statistical inference methods, such as MCMC, are likely to produce biased or even uninformative posterior estimates when multiple signals overlap due to significant interference effects. In contrast, the data-driven nature of flow-based models shows promise in yielding scientifically reliable results even under such complex conditions, as illustrated in Fig.~\ref{fig:overlap}.

Recent advances leveraging machine learning have shown significant progress in overcoming these challenges. For instance, Ref.~\cite{alvey2023things} developed a scalable framework for the joint analysis of overlapping signals using SBI. This method, implemented in the Peregrine package~\cite{Bhardwaj2023Peregrine}, is based on the TMNRE technique. Their proposed method demonstrated the ability to simultaneously characterize multiple signal sources while properly accounting for their mutual interference—an important improvement over traditional methods that typically assume isolated signals.

In principle, even tasks such as signal detection in space-based gravitational wave observation essentially boil down to parameter estimation problems. Although these applications are not the primary focus of this review, similar approaches have already seen extensive use in missions such as LISA~\cite{vílchez2024efficient} and Taiji~\cite{Ruan2023Parameter,du2024advancing,Liang2024Rapid,xu2024gravitational}.

\subsection{Population Studies}
Deep learning methods have shown significant advantages in gravitational wave population studies by enabling the efficient analysis of large event catalogs and inference of population-level parameters. Recent work has demonstrated several key advances in this area.

For example, Ref.~\cite{mould2023gravitational} combined hierarchical Bayesian modeling with a flow-based deep generative network to constrain numerical gravitational wave population models of previously intractable complexity. In this study, a 10-layer network was trained to emulate a phenomenological model with six observables and four hyperparameters, accurately and efficiently inferring the properties of a simulated catalog. This work highlights the potential of simulation-based gravitational wave population inference at unprecedented levels of complexity.

\begin{figure*}[ht!]
\centering
\includegraphics[width=0.9\textwidth]{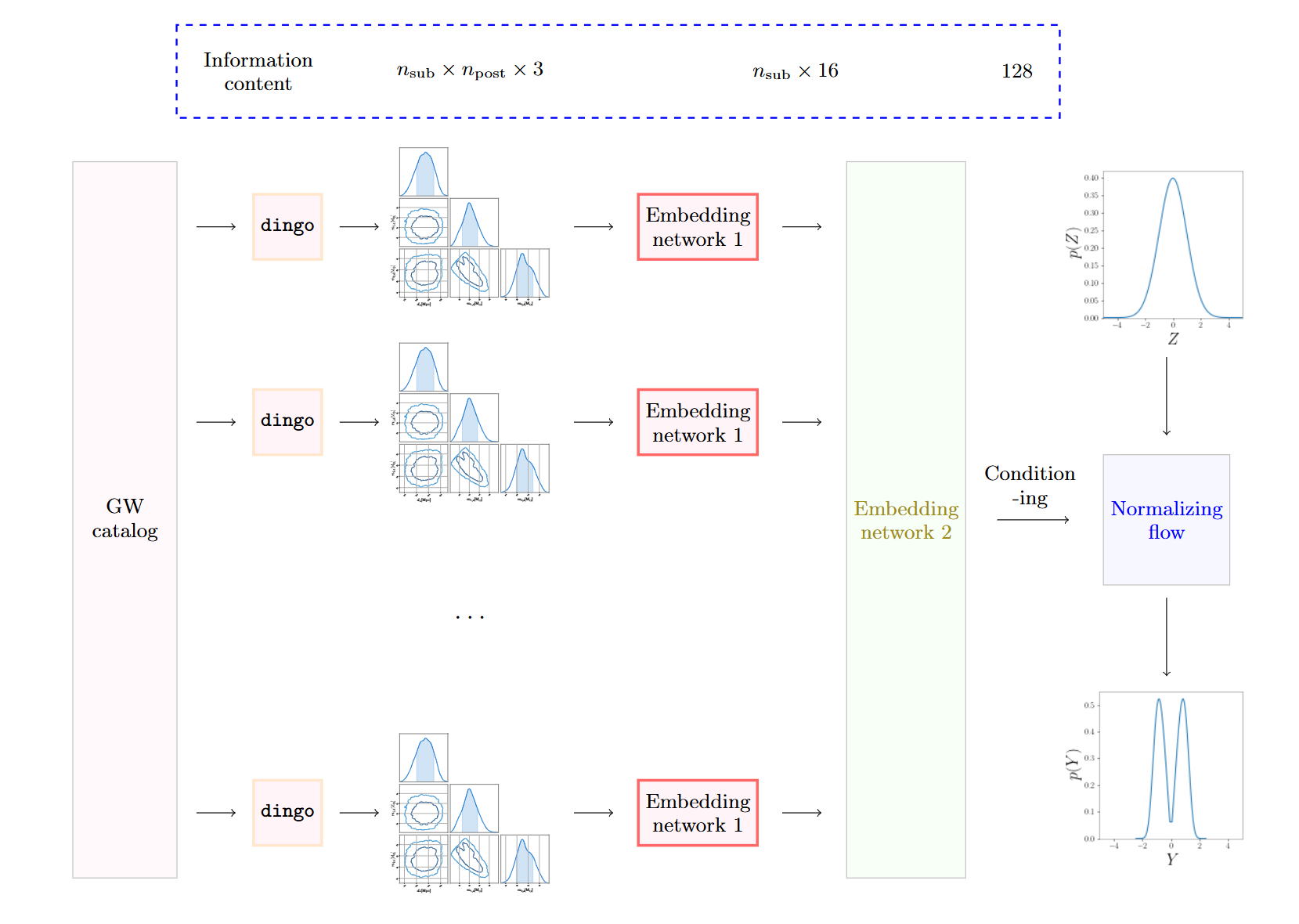}
\caption{\scriptsize
Data dimensionality reduction using embedding networks and conditioning within an NF framework. The initial data, with dimensions of \( n_{sub} \times n_{post} \times 3 \), is compressed to 128 dimensions. Notably, the embedding network remains consistent across all gravitational wave events, which streamlines the training process by reducing the number of trainable parameters. (Image adapted from Ref.~\cite{Leyde2024Gravitational})
}
\label{fig:dingo-po}
\end{figure*}

NPE techniques have proven particularly powerful for the simultaneous analysis of multiple gravitational wave events. Ref.~\cite{Leyde2024Gravitational} demonstrated how NPE can be employed for efficient population parameter inference and cosmological parameter estimation by jointly analyzing multiple events. This approach enables rapid analysis of large catalogs while appropriately accounting for measurement uncertainties in individual events, as illustrated in Fig.~\ref{fig:dingo-po}.

\subsection{Testing General Relativity and Cosmology}

One of the foremost scientific objectives of gravitational wave data analysis is to achieve scientific discovery through precise parameter estimation. In particular, tests of general relativity and the inference cosmological parameters via gravitational waves present unique challenges that modern machine-learning methods are increasingly helping to overcome. Recent work has demonstrated significant progress in both fundamental tests of general relativity and cosmological parameter estimation.

For instance, to analyze the ringdown phase of the first detected black hole merger GW150914, Ref.~\cite{Crisostomi2023Neural} developed a method using NPE with guaranteed exact coverage. Their SBI pipeline, based on masked autoregressive flows, effectively characterizes the ringdown signal. Similarly, Ref.~\cite{Pacilio2024Simulation} introduced a CNF-based simulation framework to robustly analyze the post-merger (ringdown) signal in the time domain. Their approach excels at identifying deviations from general relativity predictions while adequately accounting for uncertainties in the ringdown model.

\begin{figure}[ht!]
  \centering
  \includegraphics[width=0.5\textwidth]{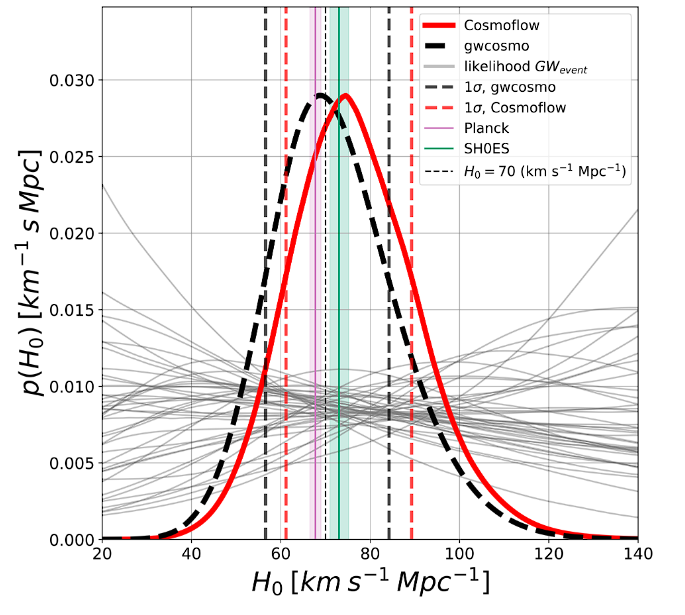}
  \caption{\scriptsize
  Combined posterior distribution of the Hubble constant \( H_0 \) estimated from 42 BBH events,  illustrating how the CosmoFlow framework combines likelihoods from multiple events to produce a robust posterior estimate under the assumption of a flat prior. (Adapted from Ref.~\cite{Stachurski2024Cosmological})
  }
  \label{fig:H0}
\end{figure}

In the realm of cosmological inference, Ref.~\cite{Stachurski2024Cosmological} developed a machine-learning approach based on NFs, commonly known as CosmoFlow, that enables the efficient and accurate estimation of cosmological parameters. This method is applicable to various types of compact binary coalescence events and cosmological models. The source code is available on GitHub~\footnote{\url{https://github.com/FedericoStachurski/CosmoFlow/}}. Fig.~\ref{fig:H0} presents the combined posterior distribution on the Hubble constant \( H_0 \) obtained by aggregating likelihoods from 42 BBH events. Assuming a flat prior, the resulting posterior demonstrates the effectiveness of the CosmoFlow framework in delivering reliable cosmological parameter estimates.

In summary, advanced machine-learning techniques—particularly those leveraging NPE and CNF frameworks—are proving to be powerful tools not only for testing the predictions of general relativity through black hole ringdown analysis but also for performing cosmological studies based on gravitational wave observations.

% \subsection{Pulsar Timing Arrays}
% % TODO  ~\cite{Shih2024Fast,vallisneri2024rapid,...}
% Pulsar timing arrays (PTAs) present unique challenges for gravitational wave detection and analysis due to their distinct observational characteristics and noise properties. Recent work has made significant progress in applying machine learning methods to address these challenges.

% The analysis of PTA data requires specialized techniques to handle:
% \begin{itemize}
%     \item Long observation timescales spanning years to decades
%     \item Complex noise models including red noise and timing noise
%     \item Sparse and irregular sampling of pulsar timing residuals
%     \item Correlations between multiple pulsars in the array
% \end{itemize}

% Machine learning approaches have shown particular promise in addressing these challenges. Recent developments include methods for:
% \begin{itemize}
%     \item Rapid characterization of gravitational wave backgrounds
%     \item Efficient handling of complex noise models
%     \item Detection and parameter estimation of individual sources
%     \item Joint analysis of multiple pulsars in the array
% \end{itemize}

% These advances are especially important as PTA sensitivity continues to improve and more candidate signals are identified. The combination of traditional methods with machine learning techniques enables more robust and efficient analysis while properly accounting for the unique characteristics of PTA observations.

%------------------------------------------------------------------------------
\section{Limitations and Future Directions}
%------------------------------------------------------------------------------
Although machine-learning approaches have shown great promise for gravitational wave data analysis, several important limitations and challenges remain to be addressed. Recent work has highlighted several key areas requiring further development.

\subsection{Limitations of Current Methods}
NPE and other SBI methods have emerged as powerful tools for gravitational wave parameter estimation~\cite{Dax2021Real,Dax2023Neural,Green2020Gravitational,Green2021Complete}.However, the following fundamental limitations must be addressed:

\begin{itemize}
    \item Model Dependence: When the underlying waveform model is updated or refined, the entire neural network requires retraining with new simulations. This process demands significant computational resources and time, potentially limiting the rapid adoption of improved waveform models in analysis pipelines.
    \item Prior Inflexibility: Neural networks inherently encode parameter priors during training, making it challenging to modify these priors post-training. This limitation restricts the exploration different prior assumptions or the updating of prior knowledge without completely retraining the network.
    \item Noise Characterization: Standard approaches typically assume Gaussian and stationary noise, which does not accurately represent real detector data. This assumption can introduce biases in the inference of key astrophysical properties, particularly when dealing with non-Gaussian noise transients (glitches).
\end{itemize}

\subsection{Possible Solutions}
In response to the limitations of current NPE and other SBI methods for gravitational wave data analysis, recent studies have proposed several potential solutions. The following subsections discuss specific solutions to the three main problems mentioned above.

\subsubsection{\textbf{Model Dependence}}

In gravitational wave data analysis, the performance of NPE is highly dependent on the model. Kolmus et al. proposed an optimization strategy to significantly improve the performance of NPE models by selecting an appropriate prior distribution and introducing a fine-tuning procedure~\cite{kolmus2024tuning}. They found that using a power law distribution as a prior for the quality parameter significantly improves sample efficiency and model performance~\cite{kolmus2024tuning}. Furthermore, after the NPE model was fine-tuned, it was able to significantly improve the sample efficiency from close to 0\% to 10\%–80\% in a short period of time~\cite{kolmus2024tuning}. These improvements not only increase the accuracy and robustness of the model but also reduce the need for computational resources, providing an efficient and accurate solution for future gravitational wave data analysis~\cite{kolmus2024tuning}.

\subsubsection{\textbf{Prior Inflexibility}}

Traditional SBI methods require fixed prior distributions, simulators, and inference tasks to be specified before training, which limits the adaptability of the models. To address this issue, Ref.~\cite{gloeckler2024allinone} introduced Simformer, a novel SBI method that combines the Transformer architecture with probabilistic diffusion models. This approach offers several advantages:

\begin{itemize}
    \item Dynamic prior adaptation during the training process, eliminating reliance on fixed prior distributions
    \item Ability to handle models with function-valued parameters and scenarios involving missing or unstructured data
    \item Demonstrated effectiveness across multiple scientific domains, suggesting potential applications in gravitational wave analysis
\end{itemize}

The emergence of Simformer represents a promising solution to the problem of inflexible priors in gravitational wave data analysis, offering enhanced flexibility without compromising inference accuracy.

\subsubsection{\textbf{Noise Characterization}}

In the context of gravitational wave analysis, the characterization of noise is crucial for accurate parameter estimation. Recent advancements have focused on leveraging machine-learning techniques to address the challenges posed by noise transients and shifts in noise distributions. The study by Raymond et al.~\cite{raymond2024simulation} presents a pioneering approach to SBI for gravitational wave from intermediate mass BBHs in real noise. This work is significant as it demonstrates the potential of using real detector noise for optimal parameter estimation using SBI methods. Moreover, some researchers~\cite{xiong2024robust,Sun_2024} developed a robust inference method for gravitational wave source parameters in the presence of noise transients using NFs. Their approach is particularly noteworthy as it provides reliable and rapid parameter inference without the need to remove glitches from the gravitational wave signal, thereby enhancing the robustness of gravitational wave data analysis. Additionally, Wildberger et al.~\cite{Wildberger2023Adapting} contributed to the field by proposing a method that adapts to noise distribution shifts in flow-based gravitational-wave inference. Their work on adapting to changes in noise characteristics is crucial for maintaining the accuracy of parameter estimations in the face of varying observational conditions. These studies collectively highlight the importance and progress in noise characterization within gravitational wave analysis, demonstrating how machine learning can be harnessed to improve the robustness and accuracy of inference methods in the presence of complex noise.

\subsection{Future Research Directions and Critical Challenges}
Although SBI methods show promise, the following fundamental challenges demand critical attention from the gravitational wave community:
\begin{itemize}
\item \textbf{Limited Interpretability:} Current SBI approaches function largely as ``black boxes", making the objective comparison of competing models difficult both theoretically and empirically. This opacity hinders scientific consensus on methodological advances.
\item \textbf{Error Attribution:} Model errors are challenging to debug and properly attribute. Distinguishing between statistical uncertainties and systematic biases remains largely unresolved, potentially undermining the scientific conclusions drawn from these methods.
\item \textbf{Robustness Concerns:} Parameter distributions from AI models often lack robustness across different noise realizations and are difficult to calibrate against established methods. Even identical methods applied to the same underlying data can produce inconsistent results when exposed to real observational data.
\item \textbf{Validation:} The lack of interpretability in AI models means that substantial additional scientific validation is required to ensure credibility and the acceptance of the results. Scientific papers using AI methods must dedicate significant space to validation procedures, comparing the methods against traditional methods and demonstrating reliability across multiple test cases.
\end{itemize}
Addressing these challenges will require interdisciplinary collaboration among gravitational wave physicists, statisticians, and machine learning experts. Future work must focus not only on developing new methods but also on establishing rigorous frameworks for model evaluation, uncertainty quantification, and scientific reliability assessment. Without these critical developments, the practical scientific impact of SBI methods in gravitational wave astronomy will remain limited despite their mathematical elegance.

%------------------------------------------------------------------------------
\section{Discussion: The Future of AI in Gravitational Wave Data Analysis}

%------------------------------------------------------------------------------
The rapid advancements in SBI and ML techniques, as outlined in this review, have undeniably transformed gravity wave data analysis. Methods such as NPE, NRE, and FMPE have demonstrated unprecedented computational efficiency, enabling real-time parameter estimation, the robust handling of overlapping signals, and large-scale population studies.

\subsection{Applicability of AI Technology in Spectrum and Source Type Analysis of Gravitational Wave Data}
The efficacy of AI-driven methods in gravitational wave astronomy is highly dependent on spectra and source type. For BNS mergers observed by ground-based detectors, the time-critical nature of multi-messenger follow-up means that parameters must be estimated within seconds after detection. The DINGO-BNS framework (Section~\ref{sec:app}) infers parameters in real time with a latency of less than a second, in contrast to traditional MCMC, which has a turnaround of several hours. This capability has proved critical in enabling the rapid identification of electromagnetic counterparts.

In the low-frequency regime (0.1 mHz–1 Hz) targeted by space-based detectors such as LISA and Taiji, extreme mass-ratio inspirals (EMRIs) create an insurmountable challenge for traditional Bayesian methods. These methods, such as MCMC, find it difficult to handle these complexities and incur high computational costs (approximately $10^{40}$ waveform evaluations), while providing finite parameter constraints that heavily rely on previous assumptions. Although the combination of template libraries and random sampling methods shows promise, it remains insufficient for exploring the full parameter space. Recent work~\cite{yun2025detection, liang2024rapidparameterestimationextreme} has shown that the use of machine-learning methods can greatly shorten the prior range of EMRIs, providing new research directions for their future scientific analysis.

However, it is important to acknowledge that for certain gravitational wave analyses, traditional methods remain entirely sufficient. When time constraints are not critical and traditional approaches are able to provide results within acceptable timeframes, the added complexity of AI methods may not be justified. Furthermore, in scenarios where Bayesian likelihood-based methods can already achieve near-optimal theoretical performance by fully modeling both the signal and noise characteristics, AI approaches may offer only marginal improvements. This is particularly true when dealing with well-understood physics, manageable parameter spaces, or when computational demands fall within reasonable limits. Nevertheless, in real-world applications, gravitational wave data often present complex noise environments and intricate signal morphologies that traditional methods struggle to model quickly and perfectly. It is precisely in these challenging scenarios—where computational efficiency, noise resilience, or model flexibility become limiting factors—that AI methods demonstrate their greatest value by overcoming specific challenges that traditional approaches cannot easily overcome.

\subsection{Notable scientific achievements achieved by SBI methods}
AI-driven SBI methods have enabled groundbreaking discoveries that traditional approaches could not have achieved. One landmark result is real-time multi-messenger astronomy: the DINGO framework (Section~\ref{sec:app}) reduced the parameter estimation latency for BNS mergers to one second, a capability unattainable using MCMC-based pipelines. Cosmologically, the CosmoFlow framework~\cite{Stachurski2024Cosmological} combined likelihoods from 42 BBH events to measure the Hubble constant independently of cosmic distance ladders.

For overlapping gravitational wave signals—a critical challenge for third-generation detectors—traditional MCMC methods not only require days of computational time but also produce biased parameter estimates. Langendorff et al.~\cite{Langendorff2022} demonstrated that NFs can perform parameter estimation on overlapped BBH systems with reasonable accuracy, solving a problem that would become practically infeasible with conventional Bayesian inference. This capability will be essential as detection rates increase and signal overlaps become commonplace in future observatories.

\subsection{The Role of AI in the Future Development of Gravitational Waves}
Although AI-driven approaches are fast and scalable, they are unlikely to entirely replace traditional Bayesian methods such as MCMC or nested sampling. Instead, the field is moving toward a hybrid paradigm in which AI and classical techniques complement each other~\cite{ICERM2025}.

AI methods transfer the computational costs to training, allowing rapid inference for individual events. However, traditional methods remain critical for validating AI-generated posteriors, especially in edge cases or scenarios requiring ultra-high precision (e.g., testing subtle deviations from general relativity). An emerging direction combines the differentiability of frameworks such as Jax with SBI techniques, as demonstrated by projects such as JaxNRSur~\cite{JaxNRSur2025}, which provides differentiable surrogate waveform models that enable gradient-based inference and optimization.

Current SBI methods suffer from rigid priors and dependency on training data. Future architectures must integrate adaptive training strategies to accommodate evolving waveform models and noise characteristics without full retraining. Techniques such as sequential inference and dynamic prior adaptation (e.g., Simformer~\cite{gloeckler2024allinone}) show promise in addressing these limitations. As detector sensitivity improves, non-Gaussian noise and overlapping signals will become more prevalent. AI methods must advance to handle these challenges autonomously, potentially by embedding physical constraints (e.g., waveform symmetries) into neural network designs or leveraging CM for robust few-shot inference.

AI will not fully supplant traditional methods but will increasingly dominate workflows in which speed and scalability are paramount. For instance, AI is already indispensable for rapid sky localization, real-time parameter estimation, and large-scale population inference. However, traditional Bayesian methods will retain their role in benchmarking, fine-grained hypothesis testing, and scenarios requiring explicit likelihood evaluations. A key question is whether these diverse techniques are mutually compatible. Projects such as flowMC~\cite{flowMC2025,Gabrie:2021tlu,Wong:2022xvh} and NESSAI~\cite{Williams_2023} demonstrate how machine learning can serve as a component within traditional MCMC and nested sampling frameworks, using NFs to accelerate sampling and improve efficiency. The future lies in synergistic pipelines in which AI accelerates the initial exploration and traditional methods refine the critical results.

In conclusion, the gravitational wave astronomy community is entering an era in which AI acts as a powerful enabler rather than a replacement. As SBI techniques mature, their integration with physical models and traditional frameworks will unlock new scientific frontiers, from probing the neutron star equation of state to resolving the Hubble tension. The challenge lies not in choosing between AI and classical methods but in harmonizing their strengths to maximize the scientific yield of current and future gravitational wave observatories.

\section{Contribution}
He Wang is supported by the National Key Research and Development Program of China (Grant No. 2021YFC2203004), the National Natural Science Foundation of China (NSFC) (Grant Nos. 12405076, 12247187, 12147103), the National Astronomical Data Center (Grant No. NADC2023YDS-01), and the Fundamental Research Funds for the Central Universities.

Throughout the development of this review, He Wang provided essential intellectual guidance. His expertise has been particularly valuable in discussing the applications of these methods across various scientific scenarios, ensuring that the review offers a balanced and comprehensive examination of the field while identifying promising future research directions. Bo Liang led the comprehensive documentation and technical analysis of SBI methods. His efforts have been instrumental in detailing the theoretical underpinnings and practical implementations of various SBI techniques, including NPE, NRE, NLE, FMPE, and CMPE, while also highlighting their relative strengths and limitations in gravitational wave data analysis.

\clearpage

\appendix

\bibliographystyle{unsrt}
\bibliography{bibtex}

\end{document}